\documentclass[a4paper,UKenglish,cleveref]{lipics-v2021}

\title{Cancellative Convex Semilattices}
\author{Ana Sokolova}{University of Salzburg, Austria}{ana.sokolova@cs.uni-salzburg.at}{https://orcid.org/0000-0002-8384-3438}{}
\author{Harald Woracek}{TU Wien, Austria}{harald.woracek@tuwien.ac.at}{https://orcid.org/0000-0002-7823-3408}{}
\authorrunning{A.~Sokolova and H.~Woracek}
\Copyright{Ana Sokolova and Harald Woracek}

\EventEditors{Corina C\^{i}rstea and Alexander Knapp}
\EventNoEds{2}
\EventLongTitle{11th Conference on Algebra and Coalgebra in Computer Science (CALCO 2025)}
\EventShortTitle{CALCO 2025}
\EventAcronym{CALCO}
\EventYear{2025}
\EventDate{June 16--18, 2025}
\EventLocation{University of Strathclyde, UK}
\EventLogo{}
\SeriesVolume{342}
\ArticleNo{12}

\usepackage{pifont}
\usepackage{tikz}
	\usetikzlibrary{arrows}
	\usetikzlibrary{patterns}
\usepackage{tikz-cd}
\usepackage{xfrac}

\newcounter{StepsCount}				
\newenvironment{Elist}{%
	\begin{list}{$\triangleright$\,Step\ \ding{\value{StepsCount}}\,:}{%
	\usecounter{StepsCount} \leftmargin=0pt \labelwidth=12pt \itemindent=\labelwidth%
	\itemsep=5pt\listparindent=\parindent} \setcounter{StepsCount}{191}}{\end{list}}

\newcommand{\Dis}[1]{${\displaystyle{#1}}$}		
\newcommand{\Dummy}{\text{\_\kern1pt}}			
\newcommand{\DS}{\mid\mkern3mu}				
\newcommand{\DP}{{\mathop:\kern5pt}}			
\newcommand{\DE}{\mathrel{\mathop:}=}			

\begin{document}

\maketitle

\begin{abstract}
	Convex semilattices are algebras that are at the same time a convex algebra and a semilattice,
	together with a distributivity axiom. These algebras have attracted some attention in the last years as suitable
	algebras for probability and nondeterminism, in particular by being the Eilenberg-Moore algebras of the nonempty
	finitely-generated convex subsets of the distributions monad. 

	\medskip
	\noindent
	A convex semilattice is cancellative if the underlying convex algebra is cancellative. Cancellative convex algebras have been characterized by M.~H.~Stone and by H.~Kneser: A convex algebra is cancellative if and only if it is isomorphic to a convex
	subset of a vector space (with canonical convex algebra operations). 

	\medskip
	\noindent
	We prove an analogous theorem for convex semilattices: A convex semilattice is cancellative if and only if it is isomorphic to a convex subset 
	of a Riesz space, i.e., a lattice-ordered vector space (with canonical convex semilattice operations). 
\end{abstract}


\section{Introduction} \label{sec:intro}

Models of computation exhibiting both nondeterministic and probabilistic behaviour have been studied for decades and are
prominent in verification~\cite{Seg95:thesis,HV98:probmiv,Baier2008,HermannsKK14,KNP02,DehnertJK017,vardi1985automatic,hansson1991time,segala1995probabilistic},
AI~\cite{CPP09,Kaelbling:1998vs,RussellNorvig:2009}, as well as in semantics~\cite{HeunenKSY17,StatonYWHK16,HermannsPSWZ11}.
Probabilities quantitatively model uncertainty and belief, whereas nondeterminism represents incomplete information, unknown
environment, implementation freedom, or concurrency. 

Algebras play an important role in the analysis of software and systems. In terms of categorical algebra,
the combination of probability and nondeterminism has been problematic, in the sense that there is no distributive law of
probability over
nondeterminism~\cite{VaraccaW06,keimel.plotkin:2017,Mio14,Jacobs08,Varacca03,Mislove00,Goubault-Larrecq08a,TixKP09a,MOW03}.
However, the relevant algebraic structures have been identified already some time ago in the above mentioned works and have
attracted additional attention recently~\cite{BSV19,BSV21,BSV22,GP20,RossetHE22}. \emph{Convex semilattices}, as coined
in~\cite{BSV19}, are such relevant algebraic structures. They are closely connected to the finitely-generated-convex-sets-of-probability-distributions monad~\cite{Mislove00,Goubault-Larrecq08a,TixKP09a, Varacca03,VaraccaW06,Jacobs08,GP20,MV20}: they are
its Eilenberg-Moore algebras as shown in~\cite{BSV19,BSV21,BSV22}. Convex semilattices also play a role in extending probability
and nondeterminism to metric spaces and quantitative equational theories~\cite{MV20,MSV21}, in semantics for higher-order
languages that combine probabilistic and nondeterministic choice~\cite{AB23}, as well as in recent work on deductive
verification~\cite{ZKST25}. In the pure mathematical context of universal algebra, convex semilattices have appeared as certain
semilattices with operators e.g.\ in \cite{hyndman.nation.nishida:2016}, or as certain modals e.g.\ in \cite{smith:2011}.

A convex semilattice is an algebra that is a convex algebra and a semilattice at the same time, including a suitable
distributivity axiom of probabilistic choice (the convex algebra operations) over nondeterminism (the semilattice operation). A
convex algebra is cancellative, if each of the convex operations, that can be understood as biased coins, is cancellative. We
call a convex semilattice cancellative if the underlying convex algebra is cancellative.

Cancellative convex algebras have been characterized by M.~H.~Stone in 1949~\cite{stone:1949} and a few years later in
1952~\cite{kneser:1952} independently by H.~Kneser: A convex algebra is cancellative if and only if it is isomorphic to a convex
subset of a vector space (with canonical convex algebra operations). 

In this paper, we prove an analogous theorem for convex semilattices: A convex semilattice is cancellative if and only if it is
isomorphic to a convex subset of a Riesz space, which is a lattice-ordered vector space (with canonical convex semilattice
operations). Interestingly, no assumption involving the semilattice operation is needed to ensure embeddability. 
	
The paper has a simple structure: In Section~\ref{sec:main}, we present the necessary background and the
main result. In Section~\ref{sec:proof} we prove our result in a sequence of steps, discussing the main ideas and tools as well
as all details. The development is elementary, with extensive use of an important geometric intuition of perspective shift that
we have already prominently employed in~\cite{sokolova.woracek:pcacon}. 
We wrap up with a short discussion in Section~\ref{sec:conc}.

\section{Background and main result} \label{sec:main}

Our main result is that cancellative convex semilattices (cf.\ \Cref{C4}) can always be embedded homomorphically into a 
Riesz space (cf.\ \Cref{C5}). 
This result is formulated in \Cref{C6} below, and is an analogue in the variety of convex semilattices 
of a classical theorem about convex algebras. To set the scene, we discuss the classical theorem first.

\subsubsection*{Revisiting convex algebras}

Abstractly, convex algebras (also known as convex sets, barycentric algebras, convex spaces, see e.g.~\cite{sokolova.woracek:pcacon} for details on related work) are the Eilenberg-Moore algebras of the finitely-supported probability distribution monad~\cite{swirszcz:1974,doberkat:2006,doberkat:2008,jacobs:2010}. Concretely, they have been studied for decades in various contexts within algebra and convex geometry. They are algebras with a family of ``biased-coin'' binary operations, used for modeling probabilistic choice. We start with recalling their definition. 

\begin{definition}
\label{C1}
	A \emph{convex algebra} $\mathbb X$ is a set $X$ together with a family of binary operations $+_p$, $p\in(0,1)$, 
	which satisfy 
	\begin{itemize}
	\item \emph{idempotence}: \Dis{\forall x\in X,p\in(0,1)\DP x+_px=x},
	\item \emph{parametric commutativity}: \Dis{\forall x,y\in X,p\in(0,1)\DP x+_py=y+_{1-p}x},
	\item \emph{parametric associativity}: 
		\[
			\forall x,y,z\in X,p,q\in(0,1)\DP (x+_py)+_qz=x+_{pq}\big(y+_{\frac{(1-p)q}{1-pq}}z\big)
			.
		\]
	\end{itemize}
	A convex algebra is called \emph{cancellative}, if it satisfies
	\begin{itemize}
	\item \Dis{\forall x,y,z\in X,p\in(0,1)\DP \big(x+_pz=y+_pz\Rightarrow x=y\big)}.
	\end{itemize}
	If it is necessary to make the operations of a convex algebra notationally explicit, 
	we write $\mathbb X=\langle X,+_p\rangle$. 
\end{definition}

\noindent
It is often practical to extend the set of operations on a convex algebra $\mathbb X=\langle X,+_p\rangle$ by defining 
\[
	x+_1y\DE x,\quad x+_0y\DE y\qquad\text{for }x,y\in X
	.
\]
Then the laws of idempotence, parametric commutativity, and parametric associativity hold even for $p,q\in[0,1]$. 

Examples of cancellative convex algebras are obtained from linear algebra.

\begin{example}
\label{C2}
	Let $V$ be a vector space over the scalar field $\mathbb R$. A subset $Y$ of $V$ is called \emph{convex}, if 
	\begin{itemize}
	\item \Dis{\forall x,y\in Y,p\in(0,1)\DP p\cdot x+(1-p)\cdot y\in Y}.
	\end{itemize}
	Let $Y$ be a convex subset of $V$, and define binary operations $+_p$ on $Y$ as 
	\begin{equation}
	\label{C18}
		x+_py\DE p\cdot x+(1-p)\cdot y\qquad\text{for }x,y\in Y,p\in(0,1)
		.
	\end{equation}
	Then $\mathbb Y\DE\langle Y,+_p\rangle$ is a cancellative convex algebra. 
\end{example}

\noindent
M.~H.~Stone~\cite{stone:1949} and H.~Kneser~\cite{kneser:1952} proved that every cancellative convex algebra (in Stone's
terminology ``barycentric algebra'' and in Kneser's terminology ``konvexer Raum'') is of 
that form. We remark that both authors work with arbitrary ordered skew fields in place of $\mathbb R$. 

\begin{theorem}[Stone-Kneser]
\label{C3}
	Let $\mathbb X$ be a cancellative convex algebra. Then there exists a vector space $V$ over $\mathbb R$ and a convex subset 
	$Y$ of $V$, such that $\mathbb X$ is isomorphic to $\mathbb Y$ (meaning the set $Y$ made into a convex algebra as in \Cref{C2}). 
\end{theorem}

\subsubsection*{Cancellative convex semilattices}

In our presentation we interchangeably work with the ``algebraic'' and the ``order theoretic'' viewpoint on semilattices.
That is: 
\begin{enumerate}
\item Algebraically, a \emph{semilattice} is a set $X$ together with a binary operation $\oplus$ that is 	idempotent, commutative, and associative, i.e., the axioms 
\begin{equation*}
\label{eq:sl-axioms}
	 x\oplus x=x, \qquad x\oplus y=y\oplus x,\qquad x\oplus (y \oplus z) = (x \oplus y) \oplus z
\end{equation*}
hold for all $x, y \in X$.
\item Order-theoretically, a \emph{semilattice} is a set $X$ together with a partial order in which each two elements have a supremum. 
\end{enumerate}
The connection between these viewpoints (going both ways) is given by 
\begin{equation}
\label{C24}
	x\leq y\mathrel{\mathop:}\Leftrightarrow x\oplus y=y,\qquad x\oplus y\DE \sup\{x,y\}
	.
\end{equation}

\begin{definition}
\label{C4}
	A \emph{convex semilattice} $\mathbb X$ is a set $X$ together with a family of binary operations $+_p$, $p\in(0,1)$, and
	another binary operation $\oplus$, such that
	\begin{itemize}
	\item $\langle X,+_p\rangle$ is a convex algebra,
	\item $\langle X,\oplus\rangle$ is a semilattice,
	\item the following distributivity law holds: 
		\[
			\forall x,y,z\in X,p\in(0,1)\DP (x\oplus y)+_pz=(x+_pz)\oplus(y+_pz)
			.
		\]
	\end{itemize}
	We call a convex semilattice \emph{cancellative}, if the convex algebra $\langle X,+_p\rangle$ is cancellative.

	If it is necessary to make the operations of a convex semilattice notationally explicit, 
	we write $\mathbb X=\langle X,+_p,\oplus\rangle$. 
\end{definition}

\noindent
Again examples of cancellative convex semilattices stem from linear algebra.

\begin{example}
\label{C5}
	An \emph{ordered vector space} is a vector space $V$ over $\mathbb R$ together with a partial order $\leq$, such that 
	\begin{itemize}
	\item \Dis{\forall x,y,z\in V\DP \big(x\leq y\Rightarrow x+z\leq y+z\big)},
	\item \Dis{\forall x\in V,q>0\DP \big(0\leq x\Rightarrow 0\leq q\cdot x\big)}.
	\end{itemize}
	An ordered vector space is called a \emph{Riesz space}, if its order is a lattice order, i.e., each two elements have a
	supremum and an infimum. 

	Let $V$ be a Riesz space, and let $Y$ be a convex subset of $V$ that is closed under binary suprema. Define binary operations $+_p$ and $\oplus$ on $Y$ as 
	\[
		x+_py\DE p\cdot x+(1-p)\cdot y\qquad\text{for }x,y\in V,p\in(0,1)
		,
	\]
	\[
		x\oplus y\DE\sup\{x,y\}\qquad\text{for }x,y\in V
		.
	\]
	Then $\mathbb Y\DE\langle Y,+_p,\oplus\rangle$ is a cancellative convex semilattice, and we refer to $\mathbb Y$ as the convex semilattice induced by $V$ on $Y$. 

	Riesz spaces go back to the work of F.~Riesz, L.~V.~Kantorovich, G.~Birkhoff and others. 
	There is a vast literature, we refer to \cite{jonge.rooij:1977,luxemburg.zaanen:1971}.
\end{example}

\noindent
We are going to prove the following theorem.

\begin{theorem}
\label{C6}
	Let $\mathbb X$ be a cancellative convex semilattice. Then there exists a Riesz space $V$ and a convex subset 
	$Y$ of $V$ that is closed under binary suprema, such that $\mathbb X$ is isomorphic to the convex semilattice induced by $V$ on $Y$. 
\end{theorem}

\begin{remark}
\label{C7}
	It may seem unnatural that we start in \Cref{C5} from a vector space ordered with a lattice order instead of
	requiring only existence of suprema as for convex semilattices. However, this is just a matter of taste, as: 
	\begin{enumerate}
	\item In every Riesz space $V$ the following law holds: 
		\begin{equation}
		\label{C8}
			\forall x,y\in V\DP \inf\{x,y\}=-\sup\{-x,-y\} .
		\end{equation}
	\item If $V$ is a convex semilattice whose underlying convex algebra is a vector space, we can use \cref{C8}
		as a definition and in this way obtain a Riesz space
		(for the sake of completeness a proof of this statement is given in \Cref{C9} below).
	\end{enumerate}
	We decided to start with a Riesz space in our presentation, since it is a classical concept and a large amount of
	theory on Riesz spaces is available.
\end{remark}

\begin{remark}
\label{C30}
	Among other algebraic structures, in \cite{keimel.plotkin:2017} also \emph{ordered barycentric algebras} are studied (see 
	Definition~2.6 in that paper). These are convex algebras that are additionally endowed with an order relation such that 
	the convex operations are monotone (they do not form an equational class of algebras). As usual, examples are obtained 
	from linear algebra: every convex subset of an ordered vector space is an ordered barycentric algebra. 
	Note that every convex semilattices in the sense of \Cref{C4} is in particular also an ordered barycentric algebra. 

	In \cite[Remarks~2.9]{keimel.plotkin:2017} it is stated (without explicit proof) that an ordered barycentric algebra can
	be embedded into an ordered vector space, if and only if the following \emph{order cancellation axiom} holds:
	\begin{equation}
	\label{C31}
		\forall x,y,z\in X,p\in(0,1)\DP \big(x+_pz\leq y+_pz\Rightarrow x\leq y\big)
		.
	\end{equation}
	Apparently, \cref{C31} implies that the underlying convex algebra is cancellative. 

	\Cref{C6} above now shows that, if suprema exist, no assumption on the order is needed 
	to guarantee embeddability into an ordered vector space (even a Riesz space). 
\end{remark}

\section{Ideas, tools, construction, proof} \label{sec:proof}

In this section we prove \Cref{C6}. The argument involves one geometric notion (Subection~\ref{subsec:p-shift}) and one rather general construction (Subsection~\ref{subsec:constr}) that together lead to the theorem.

Note already in the start, that the case when $X=\emptyset$ is trivial. Hence, from now on we assume that $X \neq \emptyset$.
We can assume without loss of generality that $X$ is a convex subset of some vector space $V$ and
that the operations $+_p$ of $X$ are inherited from $V$ as in \cref{C18}. This holds since we can embed $\langle X,+_p\rangle$ 
into some vector space by the Stone-Kneser theorem.
Moreover, we may assume that $0\in X$. This holds since we can make a translation to force that $X$
contains $0$. Note here that translations in a vector space are isomorphisms with respect to the convex algebra structure.

Our plan is to extend the semilattice operation from $X$ to some linear subspace $W$ of $V$ in such a way 
that $W$ becomes a convex semilattice, and then use \Cref{C7}{.\bf 2.} to view this as a Riesz space.

\subsection{Perspective shift}\label{subsec:p-shift}

We start with the notion of \emph{perspective shift}, that we will extensively use throughout the proof.

\begin{definition}
\label{C11}
	Let $V$ be a vector space over $\mathbb R$. Then we denote 
	\[
		P\colon\left\{
		\begin{array}{rcl}
			V\times\mathbb R\times V & \to & V
			\\
			(c,p,x) & \mapsto & px+(1-p)c
		\end{array}
		\right.
	\]
\end{definition}

\noindent
We think of $P$ as a perspective shift with center at $c$ and ratio $p$:
\\[4mm]
\rule{25mm}{0pt}
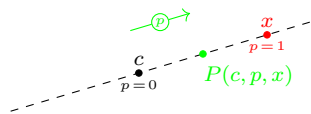
\begin{tikzpicture}[x=2pt,y=2pt,scale=0.8,font=\fontsize{8}{8}]
	\draw[dashed] (5,5)--(75,26);

	\draw[red] (65,26) node{$x$};
	\draw[red,fill=red] (65,23) circle (1.5pt);
	\draw[red] (65,20) node{\tiny$p\!=\!1$};
	
	\draw[black] (35,17) node{$c$};
	\draw[black,fill=black] (35,14) circle (1.5pt);
	\draw[black] (35,11) node{\tiny$p\!=\!0$};

	\draw[green] (48,18) node[anchor=north west]{$P(c,p,x)$};
	\draw[green,fill=green] (50,18.5) circle (1.5pt);

	\draw[green] (40,26) circle (4pt);
	\draw[green] (40,26) node{\tiny$p$};
	\draw[green,->] (42,26.6)--(47,28.1);
	\draw[green] (38,25.5)--(33,24);
\end{tikzpicture}
\\[2mm]
This function played an essential role in \cite{sokolova.woracek:pcacon}, where it was considered
on $\mathbb R^n$ (and was denoted as $\Phi_z(s,x)\DE P(z,1-s,x)$). 

Clearly, when $V$ is considered as a convex algebra in the canonical way, $P(c,p,x) = x +_p c$. 
The next several lemmas gather some properties of perspective shifts that we will need in the sequel.

\begin{lemma}
\label{C12}
	\phantom{} Let $V$ be a vector space over $\mathbb R$. Then the following properties hold:
	\begin{enumerate}
	\item \Dis{\forall c,x\in V\DP P(c,0,x)=c \text{ and } P(c,1,x)=x},
	\item \Dis{\forall c\in V,p,q\in\mathbb R\DP P(c,p,\Dummy)\circ P(c,q,\Dummy)=P(c,pq,\Dummy)},
	\item Let $c,d\in V$ and $p,q\in\mathbb R$ with $(1-p)(1-q)\neq 1$, and set 
		\begin{equation}
		\label{C22}
			s\DE\frac p{p+q-pq},\quad r\DE\frac q{p+q-pq}
			.
		\end{equation}
		Then 
		\begin{equation}
		\label{C23}
			P(d,r,\Dummy)\circ P(c,p,\Dummy)=P(c,s,\Dummy)\circ P(d,q,\Dummy)
			.
		\end{equation}
	\end{enumerate}
\end{lemma}
\begin{proof}
	The proof is carried out by simple computation.
	\begin{list}{}{\leftmargin=0pt \labelwidth=11pt \itemindent=\labelwidth \itemsep=5pt\listparindent=\parindent}
	\item[{\bf 1.}] $P(c,0,x)=0x+1c=c$, $P(c,1,x)=1x+0c=x$.
	\item[{\bf 2.}] We have 
		\begin{align*}
			P\big(c,p,P(c,q,x)\big)= &\, p\big[qx+(1-q)c\big]+(1-p)c
			\\
			= &\, pq\cdot x+\big(\underbrace{p(1-q)+(1-p)}_{1-pq}\big)\cdot c=P(c,pq,x)
			.
		\end{align*}
	\item[{\bf 3.}] The assumption that $(1-p)(1-q)\neq 1$ just says that the denominator in the definition of $r,s$ 
		is nonzero. We compute 
		\begin{align*}
			P\big(d,r,P(c,p,x)\big)= &\, r\big[px+(1-p)c\big]+(1-r)d=
			\\
			= &\, rp\cdot x+r(1-p)\cdot c+(1-r)\cdot d
			,
			\\
			P\big(c,s,P(d,q,x)\big)= &\, s\big[qx+(1-q)d\big]+(1-s)c=
			\\
			= &\, sq\cdot x+(1-s)\cdot c+s(1-q)\cdot d
			.
		\end{align*}
		Plugging the definitions of $s,r$ yields 
		\begin{align*}
			& rp=\frac{qp}{p+q-pq}=sq
			,
			\\
			& r(1-p)=\frac{q-pq}{p+q-pq}=1-\frac p{p+q-pq}=1-s
			,
			\\
			& 1-r=1-\frac q{p+q-pq}=\frac{p-pq}{p+q-pq}=s(1-q)
			.
		\end{align*}
	\end{list}
\end{proof}

As a consequence from \Cref{C12}, we immediately see that $P(c,p,\Dummy)$ is bijective with 
$P(c,p,\Dummy)^{-1} = P(c, \frac{1}{p},\Dummy)$ for all $c \in V$ and $p \in \mathbb R, p \neq 0$.  

We need one more property, a kind of associativity of perspectives.

\begin{lemma}
\label{C13}
	Let $V$ be a vector space over $\mathbb R$. Let $p,q,r,s\in\mathbb R$ be a solution of the system of equations 
	\begin{align}
		\label{C14}
		& qp=s
		\\
		\label{C15}
		& q(1-p)=(1-s)(1-r)
		\\
		\label{C16}
		& 1-q=(1-s)r
	\end{align}
	Then 
	\[
		\forall c,d,x\in V\DP P\big(d,q,P(c,p,x)\big)=P\big(P(c,r,d),s,x\big)
	\]
	\rule{25mm}{0pt}
	\begin{tikzpicture}[x=2pt,y=2pt,scale=1.2,font=\fontsize{8}{8}]
		\draw (10,10)--(40,10);
		\draw (10,10)--(10,30);
		\draw[dashed] (10,20)--(40,10);
		\draw[dashed] (10,30)--(30,10);

		\draw[black,fill=black] (10,10) circle (1.5pt);
		\draw[black] (7,7) node {$c$};
		\draw[black,fill=black] (40,10) circle (1.5pt);
		\draw[black] (42,7) node {$d$};
		\draw[black,fill=black] (10,30) circle (1.5pt);
		\draw[black] (6,32) node {$x$};
		\draw[red,fill=red] (10,20) circle (1.5pt);
		\draw[red] (9,20) node[anchor=east]{$P(c,p,x)$};
		\draw[red,fill=red] (30,10) circle (1.5pt);
		\draw[red] (30,9) node[anchor=north]{$P(c,r,d)$};
		\draw[green,fill=green] (25,15) circle (1.5pt);

		\draw[green] (23,21) circle (3pt);
		\draw[green] (23,21) node{\tiny$s$};
		\draw[green,->] (22,22)--(20,24);
		\draw[green] (24,20)--(25.5,18.5);
		\draw[green] (20,14) circle (3pt);
		\draw[green] (20,14) node{\tiny$q$};
		\draw[green,->] (18.5,14.5)--(15.5,15.5);
		\draw[green] (21.5,13.5)--(23.5,13);
	\end{tikzpicture}
\end{lemma}
\begin{proof}
	We compute 
	\begin{align*}
		P\big(d,q,P(c,p,x)\big)= &\, q\big[px+(1-p)c\big]+(1-q)d
		\\
		= &\, qp\cdot x+q(1-p)\cdot c+(1-q)\cdot d
		,
		\\
		P\big(P(c,r,d),s,x\big)= &\, sx+(1-s)\big[rd+(1-r)c\big]
		\\
		= &\, s\cdot x+(1-s)(1-r)\cdot c+(1-s)r\cdot d
		.
	\end{align*}
\end{proof}

\begin{remark}
\label{C17}
	\phantom{}
	\begin{enumerate}
	\item Any two of the three equations \cref{C14}--\cref{C16} implies the third. 
	\item The system \cref{C14}--\cref{C16} can be solved for different variables. We use:
		\begin{itemize}
		\item Given $p,q\in\mathbb R$ with $pq\neq 1$, set 
			\[
				s\DE qp,\quad r\DE\frac{1-q}{1-qp}
				.
			\]
			Then $p,q,r,s$ satisfy \cref{C14}--\cref{C16}.
		\item Given $p,r\in\mathbb R$ with $pr\neq 1$, set 
			\[
				q\DE\frac{1-r}{1-pr},\quad s\DE p\frac{1-r}{1-pr}
				.
			\]
			Then $p,q,r,s$ satisfy \cref{C14}--\cref{C16}.
		\end{itemize}
	\item If in the previous item $p,q\in(0,1)$ (or $p,r\in(0,1)$), then also $r,s\in(0,1)$ (or $q,s\in(0,1)$,
		respectively).
	\end{enumerate}
\end{remark}

We end this tools-section with one more observation about perspective shift. 
\begin{lemma}
\label{C120}
	Let $X$ be a convex semilattice that is a convex subset of a vector space $V$. Let $c \in X$ and $p \in [0,1]$. Then the perspective shift $P(c,p,\Dummy)|_X$ is a convex semilattice homomorphism.  
\end{lemma}

\begin{proof}
Let $x, y \in X$. Then 
\begin{eqnarray*}
P(c,p,x) +_q P(c,p,y) & = & (x +_p c) +_q (y +_p c)\\
& = & q(px + (1-p)c) + (1-q)(py + (1-p)c)\\
& = & p(qx + (1-q)y) + (1-p)c\\
& = & (x +_q y) +_p c\\
& = & P(c,p, x +_q y)	
\end{eqnarray*}
for any $q \in [0,1]$ and, using distributivity, 
\begin{eqnarray*}
P(c,p,x) \oplus P(c,p,y) & = & (x +_p c) \oplus (y +_p c)\\
& = & (x \oplus y) +_p c\\
& = & P(c,p, x \oplus y).
\end{eqnarray*}
\end{proof}

Of course, for arbitrary $c \in V$ and $p \in \mathbb R$, the perspective shift $P(c,p,\Dummy)$ is a linear map on $V$, 
and therefore a convex algebra homomorphism on $V$. We will also use this fact later. 

\subsection{Constructing the needed space} \label{subsec:constr}

We establish the following assertion, whose proof gives a construction of the space that we are looking for.

\begin{proposition}
\label{C10}
	Let $V$ be a vector space over $\mathbb R$ and let $X$ be a convex subset of $V$ with $0\in X$. 
	Assume that we have a binary
	operation $\oplus$ on $X$, such that $\langle X,+_p,\oplus\rangle$ is a convex semilattice (as usual $+_p$ is 
	inherited from $V$ as in \cref{C18}). Then there exists a linear subspace $W$ of $V$ with $X\subseteq W$, 
	and a binary operation $\boxplus$ on $W$, such that $\langle W,+_p,\boxplus\rangle$ (again $+_p$ as in \cref{C18}) 
	is a convex semilattice and $\boxplus|_{X\times X}=\oplus$.
\end{proposition}
\begin{proof}
	Our candidate for the subspace $W$ is 
	\[
		W\DE\big\{x\in V\DS \exists c\in X,p\in(0,1]\DP P(c,p,x)\in X\big\}
		.
	\]	
	The main idea is that we can define convex semilatttice operations on $W$ as shown in the following picture
	
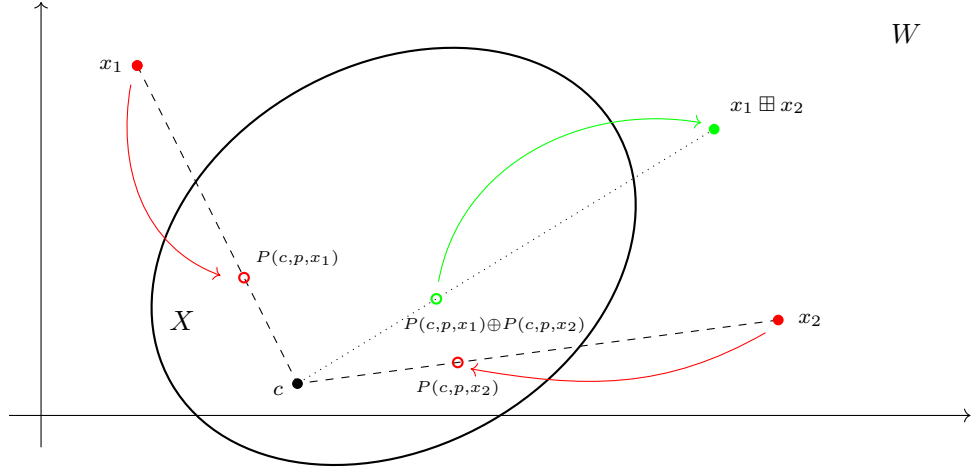
\begin{center}
\begin{tikzpicture}[x=2pt,y=2pt,scale=1.2,font=\fontsize{8}{8}]
	\draw[->] (0,5)--(150,5);
	\draw[->] (5,0)--(5,70);
	\draw (140,65) node {\large$W$};
	\draw (27,20) node {\large$X$};

	\draw[thick] (60,30) circle [x radius=40, y radius=30, rotate=30];

	\draw[dashed] (20,60)--(45,10);
	\draw[red,->] (19,57) to[out=-100,in=160] (33,27);
	\draw[dashed] (120,20)--(45,10);
	\draw[red,->] (118,18) to[out=-150,in=-10] (72,12.33);
	\draw[dotted] (110,50)--(45,10);
	\draw[green,->] (67,26) to[out=80,in=170] (108,51);

	\draw[red,fill=red] (20,60) circle (1.5pt);
	\draw[black] (16,60) node {$x_1$};
	\draw[red,fill=red] (120,20) circle (1.5pt);
	\draw[black] (125,20) node {$x_2$};
	\draw[black,fill=black] (45,10) circle (1.5pt);
	\draw[black] (42,9) node {$c$};
	\draw[red,fill=red] (20,60) circle (1.5pt);
	\draw[green,fill=green] (110,50) circle (1.5pt);
	\draw[black] (111,51) node[anchor=south west] {$x_1\boxplus x_2$};

	\draw[red,thick] (36.66,26.66) circle (1.5pt);
	\draw[black] (37,27) node[anchor=south west] {$\scriptstyle P(c,p,x_1)$};
	\draw[green,thick] (66.66,23.33) circle (1.5pt);
	\draw[black] (60,22) node[anchor=north west] {$\scriptstyle P(c,p,x_1)\oplus P(c,p,x_2)$};
	\draw[red,thick] (70,13.33) circle (1.5pt);
	\draw[black] (70,12) node[anchor=north] {$\scriptstyle P(c,p,x_2)$};
\end{tikzpicture}
\end{center}

	The fact that $X\subseteq W$ is clear: given $x\in X$ just use $c = x$ and $p = 1$. 

	We structure the proof of the required properties in five steps. Steps~\ding{192} and \ding{193} are preparatory, 
	in Step~\ding{194} we show that $W$ is a linear subspace, in Step~\ding{195} we define $\boxplus$, and in 
	Step~\ding{196} we prove that $W$ becomes a convex semilattice.
\begin{Elist}
\item
	Let $x\in W$ and let $c\in X,p\in(0,1]$ be such that $P(c,p,x)\in X$. Set 
	\[
		p(x)\DE\sup\big\{p\in(0,1]\DS P(c,p,x)\in X\big\}
		.
	\]
	We show that $P(c,q,x)\in X$ for all $q\in(0,p(x))$. 
	To see this, assume $q\in(0,p(x))$, and choose $p\in[q,p(x)]$ with $P(c,p,x)\in X$. 
	Then $\frac qp\in[0,1]$, and hence, by \Cref{C12}{.\bf 2.}, 
	\[
		P(c,q,x)=P\Big(c,\frac qp,P(c,p,x)\Big)\in X
		.
	\]
\item
	Let $x_1,\ldots,x_n\in W$. We show that there exists $c\in X,p\in(0,1]$, such that $P(c,p,x_i)\in X$ for all 
	$i\in\{1,\ldots,n\}$. 

	To start with, consider $x\in W$, let $c\in X,p\in(0,1]$ with $P(c,p,x)\in X$, and let $d\in X,r\in(0,1)$. 
	Then, by \Cref{C13} and \Cref{C17}{.\bf 2.}, we have 
	\begin{equation}
	\label{C20}
		P\Big(P(c,r,d),\underbrace{p\frac{1-r}{1-pr}}_{\in(0,1]},x\Big)
		=P\Big(d,\underbrace{\frac{1-r}{1-pr}}_{\in(0,1]}P(c,p,x)\Big)\in X
		.
	\end{equation}
	Now consider $x_1,\ldots,x_n\in W$. Choose $c_i\in X,p_i\in(0,1]$ with $P(c_i,p_i,x_i)\in X$. Set 
	\[
		c\DE\sum_{i=1}^n\frac 1nc_i
		,
	\]
	then $c\in X$ as $X$ is a convex algebra and hence it is also closed under arbitrary convex combinations. We have 
	\[
		c=P\Big(c_i,\frac{n-1}n,\sum_{\substack{j=1\\ j\neq i}}^n\frac 1{n-1}c_j\Big)
		,
	\]
	and by \cref{C20} thus 
	\[
		P\Big(c,p_i\frac{1-\frac{n-1}n}{1-p_i\frac{n-1}n},x_i\Big)
		=P\Big(\sum_{\substack{j=1\\ j\neq i}}^n\frac 1{n-1}c_j,\frac{1-\frac{n-1}n}{1-p_i\frac{n-1}n},P(c_i,p_i,x_i)\Big)
		\in X
		.
	\]
	Now choose $p$ with $0<p<\min\limits_{i\in\{1,\ldots,n\}}p_i\frac{1-\frac{n-1}n}{1-p_i\frac{n-1}n}$ and use 
	Step~\ding{192}: it follows that indeed $P(c,p,x_i) \in X$.
\item
	We show that $W$ is a linear subspace of $V$ and contains $X$. It is easy to see that $W$ is convex. 
	Given $x,y\in W$ choose $c\in X$ and 
	$p\in(0,1]$ with $P(c,p,x)\in X$ and $P(c,p,y)\in X$, which is possible by Step~\ding{193}. Then, for each $q\in[0,1]$, 
	\begin{align*}
		P(c,p,qx+(1-q)y)= &\, p\big[qx+(1-q)y\big]+(1-p)c
		\\
		= &\, q\big[px+(1-p)c\big]+(1-q)\big[py+(1-p)c\big]\in X
		,
	\end{align*}
	and hence $qx+(1-q)y\in W$. 

	Next we show that $x\in W$ and $q\geq 0$ implies $qx\in W$. If $q\in[0,1]$ we can use that $0\in X\subseteq W$ and 
	$W$ is already known to be convex to arrive at $qx=qx+(1-q)0\in W$. Assume now that $q>1$. Choose $c\in X,p\in(0,1]$ 
	with $P(c,p,x)\in X$. We apply \Cref{C13} and \Cref{C17}{.\bf 2.} to obtain, with $p$ in place of $q$, $\frac{1}{q}$ in place of $p$,  $qx$ in place of $x$, and $c, d$ exchanged:
	\\[0mm]
	\rule{10mm}{0pt}
	\begin{tikzpicture}[x=2pt,y=2pt,scale=1.2,font=\fontsize{8}{8}]
		\draw (10,10)--(40,10);
		\draw (10,10)--(10,30);
		\draw[dashed] (10,20)--(40,10);
		\draw[dashed] (10,30)--(30,10);

		\draw[black,fill=black] (10,10) circle (1.5pt);
		\draw[black] (7,7) node {$0$};
		\draw[black,fill=black] (40,10) circle (1.5pt);
		\draw[black] (42,7) node {$c$};
		\draw[black,fill=black] (10,30) circle (1.5pt);
		\draw[black] (6,32) node {$qx$};
		\draw[red,fill=red] (10,20) circle (1.5pt);
		\draw[red] (7,20) node {$x$};
		\draw[red,fill=red] (30,10) circle (1.5pt);
		\draw[green,fill=green] (25,15) circle (1.5pt);
		\draw[green] (25,14) node[anchor=south west]{$P(c,p,x)$};
	\end{tikzpicture}
	\raisebox{5pt}{
	\begin{minipage}[b]{70mm}
		\small
		\begin{multline*}
			P\Big(\underbrace{P\Big(0,\frac{1-p}{1-\frac pq},c\Big)}_{\in X},\underbrace{\frac pq}_{\in[0,1)},qx\Big)
			\\
			=P\big(c,p,\underbrace{P\big(0,\tfrac 1q,qx\big)}_{\in X}\big)\in X
		\end{multline*}
	\end{minipage}
	}
	\\[2mm]
	and hence $qx \in W$. 
	
	Finally, we show that $x\in W$ implies $-x\in W$. Choose $c\in X,p\in(0,1)$ with $P(c,p,x)\in X$. 
	Avoiding $p=1$ is possible by Step~\ding{192}.  
	Then again using \Cref{C13} (with $c = x$ and $d = c$) and the second part of \Cref{C17}{.\bf 2.} by simultaneously substituting $\frac 12$ for $p$ 
	and $1-p$ for $r$, we obtain $q$ and $s$, which are in $(0,1)$ by \Cref{C17}{.\bf 3.}, with 
	\\[0mm]
	\rule{10mm}{0pt}
	\begin{tikzpicture}[x=2pt,y=2pt,scale=1.2,font=\fontsize{8}{8}]
		\draw (10,10)--(40,10);
		\draw (10,10)--(10,30);
		\draw[dashed] (10,20)--(40,10);
		\draw[dashed] (10,30)--(30,10);

		\draw[black,fill=black] (10,10) circle (1.5pt);
		\draw[black] (7,7) node {$x$};
		\draw[black,fill=black] (40,10) circle (1.5pt);
		\draw[black] (42,7) node {$c$};
		\draw[black,fill=black] (10,30) circle (1.5pt);
		\draw[black] (6,32) node {$-x$};
		\draw[red,fill=red] (10,20) circle (1.5pt);
		\draw[red] (7,20) node {$0$};
		\draw[red,fill=red] (30,10) circle (1.5pt);
		\draw[green,fill=green] (25,15) circle (1.5pt);
		\draw[green] (25,14) node[anchor=south west]{$P\big(c,\tfrac p{\raisebox{-1.5pt}{\sfrac 12}+p},x\big)$};
	\end{tikzpicture}
	\raisebox{5pt}{
	\begin{minipage}[b]{70mm}
		\small
		\begin{multline*}
			P\Big(\underbrace{P(x, 1-p,c)}_{=P(c,p,x)\in X},
			\underbrace{s}_{\in[0,1)},-x\Big)
			\\
			=P\Big(c,q,\underbrace{P(x,\tfrac 12,-x)}_{=0}\Big) \in X
		\end{multline*}
	\end{minipage}
	}
	
	\noindent and thus $-x$ satisfies the defining property of $W$. The numerical values of these $q$ and $s$ are $\frac{2p}{1+p}$ and $\frac{p}{1+p}$, respectively.
\item
	We show that by the following procedure a binary operation on $W$ extending $\oplus$ is well-defined: 
	given $x_1,x_2\in W$, choose $c\in X,p\in(0,1]$ with $P(c,p,x_i)\in X$, $i\in\{1,2\}$, and set
	\begin{equation}
	\label{C21}
		x_1\boxplus x_2\DE P\Big(c,\frac 1p,P(c,p,x_1)\oplus P(c,p,x_2)\Big)
		.
	\end{equation}
	For the time being, denote the element of $V$ given by the right hand side of \cref{C21} as $w$. First, 
	\[
		P(c,p,w)=P(c,p,x_1)\oplus P(c,p,x_2)\in X
		,
	\]
	and it follows that $w\in W$. Now assume we have $d\in X,q\in(0,1]$ that also satisfy $P(d,q,x_i)\in X$, $i=1,2$. Since 
	$p,q\in(0,1]$, we have $(1-p)(1-q)\neq 1$ and \Cref{C12}{.\bf 3.} applies. Taking inverses in \cref{C23} (here $r,s$ are 
	as in \cref{C22} and we note that $r,s\in(0,1]$) yields that also 
	\[
		P(c,\tfrac 1p,\Dummy)\circ P(d,\tfrac 1r,\Dummy)=P(d,\tfrac 1q,\Dummy)\circ P(c,\tfrac 1s,\Dummy)
		.
	\]
	Set 
	\[
		y_i\DE P\big(d,r,\underbrace{P(c,p,x_i)}_{\in X}\big)=P\big(c,s,\underbrace{P(d,q,x_i)}_{\in X}\big)\in X
		.
	\]
	The maps $P(d,r,\Dummy)|_X$ and $P(c,s,\Dummy)|_X$ are convex semilattice homomorphisms on $X$, by \Cref{C120}, and hence 
	\[
		y_1\oplus y_2=P\big(d,r,P(c,p,x_1)\oplus P(c,p,x_2)\big)=P\big(c,s,P(d,q,x_1)\oplus P(d,q,x_2)\big)
		.
	\]
	We obtain 
	\begin{eqnarray*}
		P\big(c,\tfrac 1p,P(c,p,x_1)\oplus P(c,p,x_2)\big)
		& = & \big[P(c,\tfrac 1p,\Dummy)\circ P(d,\tfrac 1r,\Dummy)\big](y_1\oplus y_2)\\
		& = &\big[P(d,\tfrac 1q,\Dummy)\circ P(c,\tfrac 1s,\Dummy)\big](y_1\oplus y_2)\\
		& = & P\big(d,\tfrac 1q,P(d,q,x_1)\oplus P(d,q,x_2)\big).
	\end{eqnarray*}
	which proves that $w$ is independent of the choice of $c$ and $q$.

	Finally, note that for $x_1,x_2\in X$ we can choose $p=1$ (and arbitrary $c\in X$). 
	This choice gives $x_1\boxplus x_2=x_1\oplus x_2$.
\item
	We show that $W$ is a convex semilattice with $\boxplus$. We start with the semilattice axioms, which are checked simply 
	by unfolding the definitions. Let $c\in X$ and $p\in(0,1]$ always be chosen appropriately according to the definition 
	of $W$.\\ Idempotence is immediate:
	\[
	{x\boxplus x=P\big(c,\tfrac 1p,P(c,p,x)\oplus P(c,p,x)\big)=P\big(c,\tfrac 1p,P(c,p,x)\big)=x}.
	\]
	Similarly, commutativity follows:
	\[
	{x\boxplus y=P\big(c,\tfrac 1p,P(c,p,x)\oplus P(c,p,y)\big)=P\big(c,\tfrac 1p,P(c,p,y\oplus P(c,p,x))\big)
		=y\boxplus x}
	.	\]
	Finally, we derive associativity by:
	\begin{align*}
		\mkern-27mu
			(x\boxplus y)\boxplus z
			= &\, P\big(c,\tfrac 1p,P(c,p,x\boxplus y)\oplus P(c,p,z)\big)
			\\
			= &\, P\big(c,\tfrac 1p,P\big(c,p,P\big(c,\tfrac 1p,P(c,p,x)\oplus P(c,p,y)\big)\big)\oplus P(c,p,z)\big)
			\\
			= &\, P\big(c,\tfrac 1p,\big[P(c,p,x)\oplus P(c,p,y)\big]\oplus P(c,p,z)\big)
			\\
			= &\, P\big(c,\tfrac 1p,P(c,p,x)\oplus\big[P(c,p,y)\oplus P(c,p,z)\big]\big)
			\\
			= &\, P\big(c,\tfrac 1p,P(c,p,x)\oplus P(c,p,y\boxplus z)\big)
			\\
			= &\, x\boxplus(y\boxplus z).
		\end{align*}
	The proof of distributivity is slightly more complicated. We first show a particular case, namely, 
	\begin{equation}
	\label{C25}
		\forall x,y\in W,d\in X,q\in(0,1)\DP (x\boxplus y)+_qd=(x+_qd)\boxplus(y+_qd)
	\end{equation}
	Choose $c,p$ suitable and set again $s\DE\frac p{p+q-pq},r\DE\frac q{p+q-pq}$. Then, using \cref{C23} twice, also in a somewhat peculiar way --- by composing with the inverse of $P(c,s,\Dummy)$ on the left and the inverse of $P(c,p,\Dummy)$ on the right, we get 
	\begin{align*}
		(x\boxplus y)+_qd
		= &\, P\big(d,q,P\big(c,\tfrac 1p,P(c,p,x)\oplus P(c,p,y)\big)\big)
		\\
		= &\, P\big(c,\tfrac 1s,P\big(d,r,P(c,p,x)\oplus P(c,p,y)\big)\big)
		\\
		= &\, P\big(c,\tfrac 1s,P(d,r,P(c,p,x))\oplus P(d,r,P(c,p,y))\big)
		\\
		= &\,  P\big(c,\tfrac 1s,P(c,s,P(d,q,x))\oplus P(c,s,P(d,q,y))\big)
		\\
		= &\, P(d,q,x)\boxplus P(d,q,y)
		\\
		= &\, (x+_qd)\boxplus(y+_qd).
	\end{align*}
	Note here that $P(c,s,P(d,q,x)),P(c,s,P(d,q,y))\in X$. 

	Now let $x,y,z\in W$ and $p\in(0,1)$. Choose $d\in X,q\in(0,1]$ with $P(d,q,x),P(d,q,y),$ $P(d,q,z)\in X$. 
	Using that $P(d,q,\Dummy)$ is a convex algebra homomorphism, the definition of $\boxplus$, distributivity in $X$, once
	again that $P(d,q,\Dummy)$ is a convex algebra homomorphism, the coincidence of $\boxplus$ and $\oplus$ on $X$, and
	finally \cref{C25} translated in terms of perspective shift, we find
	\begin{align*}
		P\big(d,q, &\, (x\boxplus y)+_pz\big)
		=\Big[P(d,q,x\boxplus y)\Big]+_pP(d,q,z)
		\\
		= &\, \Big[P(d,q,x)\oplus P(d,q,y)\Big]+_pP(d,q,z)
		\\
		= &\, \Big[P(d,q,x)+_pP(d,q,z)\Big]\oplus\Big[P(d,q,y)+_pP(d,q,z)\Big]
		\\
		= &\, P(d,q,x+_pz)\oplus P(d,q,y+_pz)
		\\
		= & P(d,q,x+_pz)\boxplus P(d,q,y+_pz)
		\\
		= &\, P\big(d,q,(x+_pz)\boxplus(y+_pz)\big).
	\end{align*}
	Since $P(d,q,\Dummy)$ is injective, it follows that $(x\boxplus y)+_pz=(x+_pz)\boxplus(y+_pz)$.
\end{Elist}
\end{proof}


\subsection{The constructed space is a Riesz space}

The proof of \Cref{C6} will be complete once we prove the following lemma already mentioned in \Cref{C7}.

\begin{lemma}
\label{C9}
	Let $V$ be a vector space over $\mathbb R$ and set $x+_py\DE px+(1-p)y$ for $x,y\in V$, $p\in(0,1)$. 
	Further, let $\oplus$ be a binary operation on $V$ such that $\langle V,+_p,\oplus\rangle$ is a convex semilattice, and
	let $\leq$ be the partial order induced by $\oplus$ via \cref{C24}. Then $\langle V,\leq\rangle$ is a Riesz space. 
\end{lemma}
\begin{proof}
	We first show that 
	\begin{equation}
	\label{C19}
		\forall x,y\in V,q>0\DP qx\oplus qy=q(x\oplus y)
	\end{equation}
	If $q\leq 1$, we can use the distributive law to compute 
	\[
		q(x\oplus y)=(x\oplus y)+_q0=(x+_q 0)\oplus(y+_q0)=qx\oplus qy
		.
	\]
	If $q>1$, we apply this rule with $\frac 1q$ and $qx,qy$ to obtain 
	\[
		\frac 1q\big(qx\oplus qy\big)=x\oplus y
		.
	\]
	Hence, also in this case $qx\oplus qy=q(x\oplus y)$.

	From \cref{C19} we can now deduce that $\langle V,\leq\rangle$ is an ordered vector space. First, for $x,y\in V$ and $q>0$, 
	\[
		x\leq y\Leftrightarrow x\oplus y=y\Leftrightarrow q(x\oplus y)=qy\Leftrightarrow 
		qx\oplus qy=qy\Leftrightarrow qx\leq qy	
	\]
	and in particular for $x =0$ we get first needed property for ordered space. \\
	
	\noindent Second, for $x,y,z\in V$ with $x\leq y$ we have 
	\begin{eqnarray*}
		(x+z)\oplus(y+z) & = &  2(x+_{\frac 12}z) \oplus 2(y+_{\frac 12}z)\\
		& = & 2\big[(x+_{\frac 12}z)\oplus(y+_{\frac 12}z)\big]\\
		& = &
		2\big[(x\oplus y)+_{\frac 12}z\big]=2(y+_{\frac 12}z)\\
		& = & y+z
		.
	\end{eqnarray*}
	It remains to establish existence of infima. For each two elements $x,y\in V$ it holds that 
	\[
		x\leq y\Leftrightarrow x+(-x-y)\leq y+(-x-y)\Leftrightarrow -y\leq -x
		.
	\]
	Now let $x,y\in V$ be given. Then 
	\[
		\{z\in V\DS -x\leq z\wedge -y\leq z\}=\{z\in V\DS -z\leq x\wedge -z\leq y\}
		,
	\]
	and hence $-\sup\{-x,-y\}$ is the infimum of $x$ and $y$.\\ 
\end{proof}

\section{Conclusions} \label{sec:conc}

We have proven an analogue in the setting of convex semilattices of what we refer to as the Stone-Kneser theorem for cancellative
convex algebras. We showed in \Cref{C6} that a convex semilattice is cancellative if and only if it can be homomorphically embedded 
in a Riesz space. Here, we call a convex semilattice cancellative when its underlying convex algebra is cancellative. 

Interestingly, as mentioned in \Cref{C30}, the result does not impose any condition on the semilattice structure. 

The work presented here is a first step towards understanding the structure of convex semilattices. Other directions that we will explore involve more specific notions of cancellativity as well as understanding of homomorphisms and congruences of convex semilattices. 

By more specific notions of cancellativity of convex semilattices, that are also worth investigating, we mean, e.g., the following property that we call ``mid-point cancellativity''
\[
x +_{\frac 12} y = u +_{\frac 12} v \wedge x \oplus y = u \oplus v \Rightarrow (x = u \wedge y = v) \vee (x = v \wedge y = u) .
\]
We have some preliminary results on this topic that we will fully explore in future work.


\bibliography{CSL-CAN-lipics-v2021}

\end{document}